\documentclass[11pt]{article}

\usepackage[margin=1in]{geometry}
\usepackage{graphicx}
\usepackage{amsmath}
\usepackage{amssymb}
\usepackage{booktabs}
\usepackage{array}
\usepackage{microtype}
\usepackage{cite}
\usepackage{longtable}
\usepackage[hidelinks]{hyperref}

\graphicspath{{figures/}}

\title{\textbf{Distilling Lexical Product Associations into Deep Transformers:\\
An Extreme Multi-Label Approach for Natural Language E-Commerce Search}}

\author{
  \textbf{Sunnidhya Roy}\thanks{Equal contribution.}\\
  Department of Data Science and Artificial Intelligence\\
  International Institute of Information Technology Bangalore\\
  Bengaluru, India\\
  \texttt{rsunnidhya@gmail.com}
  \and
  \textbf{Samarpita Bhaumik}\footnotemark[1]\\
  Department of Data Science and Artificial Intelligence\\
  International Institute of Information Technology Bangalore\\
  Bengaluru, India\\
  \texttt{samarpitabhaumik2017@gmail.com}
}

\date{September 2026\\[1ex]
\textbf{Link}: \href{https://github.com/Sunnidhya/Distilling-Lexical-Product-Associations-into-Deep-Transformers}{\textbf{GitHub}}}

\begin{document}

\maketitle

\begin{abstract}
Traditional e-commerce search platforms rely heavily on inverted indices and token-level lexical matching algorithms (e.g., BM25 and TF-IDF), which frequently fail when faced with conversational, intent-driven, or paraphrased user queries---the classic \emph{vocabulary mismatch} problem. In this work, we formulate conversational product recommendation as an \textbf{Extreme Multi-Label Classification (XMLC)} problem over an e-commerce catalog of $N = 54{,}000$ products spanning 27 balanced retail categories from the \textit{Amazon Reviews '23} benchmark. Using a pre-trained DistilBERT transformer encoder, we distill dense item-to-item similarity topologies (generated via TF-IDF cosine similarity over cumulative metadata with $K = 50$ nearest neighbours) into a deep contextual representation via a \emph{pseudo-label knowledge distillation} framework.

Evaluated on an exact $85/15$ train/validation split ($8{,}089$ held-out products across $C = 53{,}923$ extreme multi-label output classes) with strict self-exclusion enforced throughout, the DistilBERT neural student achieves $P@1 = 93.15\%$, $P@5 = 90.08\%$, $\text{NDCG}@10 = 0.8845$, and $\text{MRR}@10 = 0.9545$, closely recovering the empirical ceiling established by the corrected TF-IDF teacher ($P@1 = 98.10\%$, $\text{NDCG}@10 = 0.9419$, $\text{MRR}@10 = 0.9882$). Furthermore, a qualitative benchmark across ten structured natural language query archetypes---encompassing situational, cross-category, paraphrased, and negative-constraint queries---demonstrates that the transformer student generalises substantially beyond keyword matching, successfully resolving implicit user intent where lexical models fail completely. Finally, we analyse the architectural and memory scalability trade-offs of extreme classification projection layers at industrial catalog scale ($>10^6$ items) and present a concrete deployment trajectory toward Dual-Encoder (Two-Tower) vector search.
\end{abstract}

\section{Introduction}

In the contemporary digital commerce ecosystem, recommendation and search systems serve as essential infrastructure guiding product discovery, customer conversion, and user satisfaction~\cite{covington2016}. Traditional industrial e-commerce retrieval architectures predominantly rely on three conventional paradigms:
\begin{itemize}
  \item \textbf{Collaborative Filtering (CF):} Operates on user-item interaction histories (e.g., matrix factorization or item-based collaborative filtering)~\cite{koren2009,sarwar2001}. While effective for active users, CF suffers acutely from the \emph{cold-start problem}~\cite{schein2002} when encountering newly introduced products or anonymous visitors lacking historical click/purchase records.
  \item \textbf{Content-Based Filtering (CBF):} Analyzes structured product metadata and catalog attributes. However, conventional CBF often induces \emph{filter bubbles} and catalog over-specialization, repeatedly recommending nearly identical variants within a single subcategory.
  \item \textbf{Lexical Inverted-Index Engines:} Systems relying on TF-IDF or BM25 enforce rigid token-matching constraints. When users pose conversational queries (e.g., \textit{``I have a birthday party tomorrow, suggest me some products''}), inverted indices return irrelevant results or empty result sets because the literal tokens \textit{``birthday''}, \textit{``party''}, and \textit{``tomorrow''} rarely appear verbatim inside the technical specifications of banners, cake toppers, or festive party supplies.
\end{itemize}

This linguistic discrepancy between casual user expressions and formal product descriptions is known as the \emph{vocabulary mismatch} problem~\cite{furnas1987,tfidf,bm25}. Addressing this gap requires transitioning from syntactic keyword matching to deep \textbf{semantic retrieval}.

In this paper, we present an end-to-end framework that reframes multi-category e-commerce recommendation as an \textbf{Extreme Multi-Label Classification (XMLC)} distillation task~\cite{fastxml,xmlc,hinton2015}. Specifically, we leverage an unsupervised lexical teacher (TF-IDF with $K = 50$ nearest neighbours) to construct a pseudo-ground-truth association graph across $54{,}000$ products from 27 balanced categories of the \textit{Amazon Reviews '23} benchmark~\cite{amazon}. We then train a DistilBERT transformer student~\cite{distilbert} to map raw, unstructured textual metadata into the $53{,}923$-dimensional label space using multi-label Binary Cross-Entropy with Logits.

The primary contributions of this paper are fourfold:
\begin{enumerate}
  \item \textbf{Principled Distillation Framing:} We formalize the transfer of sparse lexical item topologies into a compact, bidirectional transformer representation through pseudo-label learning, resolving prior circular evaluation pitfalls.
  \item \textbf{Resolution of Baseline Alignment \& Rigorous Evaluation:} We identify and formally rectify an index-misalignment bug present in standard baseline implementations and conduct rigorous evaluation with strict self-exclusion ($-\infty$ masking on self-labels) across five Information Retrieval metrics ($P@k$, $R@k$, $\text{NDCG}@k$, $\text{MRR}@k$, and $\text{MAP}@k$ for $k \in \{1, 3, 5, 10, 20\}$).
  \item \textbf{Qualitative Generalization Benchmarking:} Through a curated test suite of ten diverse natural language query archetypes, we demonstrate empirical evidence of intent recovery and semantic comprehension where keyword matching degenerates.
  \item \textbf{Architectural Scalability Analysis:} We provide a detailed technical analysis of the memory and computational bottlenecks inherent in extreme multi-label classification projection layers ($C \approx 54{,}000$ vs.\ $C > 10^6$) and delineate a concrete roadmap toward Dual-Encoder (Two-Tower) Approximate Nearest Neighbor (ANN) vector search.
\end{enumerate}

\section{Related Work}\label{sec:related}

\paragraph{Lexical Retrieval and Search Engines.}
Classical text retrieval systems rely on term-weighting algorithms such as TF-IDF~\cite{tfidf} and the probabilistic BM25 ranking framework~\cite{bm25,manning2008}. While computationally lightweight and scalable via inverted index data structures, these methods are fundamentally limited by exact token matching and cannot capture synonyms, conceptual relationships, or syntactic paraphrasing.

\paragraph{Pre-Trained Language Models for Dense Retrieval.}
The emergence of pre-trained transformers, notably BERT~\cite{bert}, established contextualized token representations as the standard foundation for natural language understanding. Sanh et al.~\cite{distilbert} introduced DistilBERT, compressing the BERT-base architecture by $40\%$ while retaining $97\%$ of its downstream language comprehension capabilities and running $60\%$ faster. In information retrieval, Dense Passage Retrieval (DPR)~\cite{dpr} established that dual-encoder transformers projecting queries and documents into a shared latent space dramatically outperform BM25 in open-domain question answering.

\paragraph{Knowledge Distillation and Pseudo-Labeling.}
Knowledge distillation, pioneered by Hinton et al.~\cite{hinton2015}, enables the transfer of knowledge from a high-capacity teacher to a resource-efficient student model. In semi-supervised and self-supervised paradigms, pseudo-labeling leverages automated heuristics or unsupervised graph clustering to generate supervisory signals in domains where human annotations are absent or prohibitively costly. In our work, we adapt this paradigm by utilizing an unsupervised lexical graph as the teacher to supervise a deep neural sequence encoder.

\paragraph{Extreme Multi-Label Classification (XMLC).}
XMLC involves assigning relevant subset labels from a discrete pool containing tens of thousands to millions of candidate categories~\cite{xmlc}. Modern XMLC approaches commonly utilize deep transformer backbones to handle fine-grained text semantics~\cite{attentionxml}. Our formulation models product recommendation directly as XMLC over $C = 53{,}923$ items, providing an explicit testbed for evaluating transformer generalization under extreme label sparsity.

\section{Problem Formulation and Methodology}\label{sec:theory}

\subsection{Mathematical Task Definition}
Let $\mathcal{C} = \{p_1, p_2, \dots, p_N\}$ denote an e-commerce catalog comprising $N = 54{,}000$ distinct products distributed evenly across $|\mathcal{K}| = 27$ categories ($2{,}000$ products per category). Each product $p_i$ is characterized by a cumulative textual representation $x_i \in \mathcal{X}$, formed by concatenating its title, detailed description, and primary retail category:
\begin{equation}
  x_i = \operatorname{Concat}\left(\texttt{title}_i, \; \texttt{description}_i, \; \texttt{category}_i\right).
\end{equation}
Each product in the catalog is assigned a unique integer identifier $c \in \{0, 1, \dots, C-1\}$, where $C = 53{,}923$ denotes the total number of unique classes retained after preprocessing.

\subsection{Lexical Teacher \& Pseudo-Label Construction}
To construct pseudo-labels without expensive human annotation, we establish an unsupervised lexical baseline. For each item $p_i$, we extract its $L_2$-normalized TF-IDF feature vector $\mathbf{t}_i \in \mathbb{R}^{|V|}$ over a vocabulary $|V| = 50{,}000$ using sublinear term-frequency scaling ($\text{tf}_{\text{scaled}} = 1 + \log(\text{tf})$). Pairwise lexical similarity between products $p_i$ and $p_j$ is defined by cosine similarity:
\begin{equation}
  S_{ij} = \cos(\mathbf{t}_i, \mathbf{t}_j) = \frac{\mathbf{t}_i \cdot \mathbf{t}_j}{\|\mathbf{t}_i\|_2 \|\mathbf{t}_j\|_2}.
\end{equation}
For every product $p_i$, we identify its $K = 50$ nearest neighbours within the corpus, explicitly excluding the item itself ($j \neq i$):
\begin{equation}
  \mathcal{N}_K(i) = \operatorname{argtop}_K \left( \{ S_{ij} \mid j \in \{1, \dots, N\} \setminus \{i\} \} \right).
\end{equation}
The ground-truth multi-hot target vector $\mathbf{y}_i \in \{0, 1\}^C$ for product $p_i$ is defined as:
\begin{equation}
  y_{i,c} = \begin{cases}
    1, & \text{if } \exists\, p_j \in \mathcal{N}_K(i) \text{ such that } \operatorname{class}(p_j) = c, \\
    0, & \text{otherwise.}
  \end{cases}
\end{equation}
Because each item has approximately 50 positive labels out of 53,923 classes, the label density is extremely sparse:
\begin{equation}
  \rho = \frac{K}{C} \approx \frac{50}{53{,}923} \approx 9.27 \times 10^{-4} \quad (0.093\%).
\end{equation}
To prevent memory exhaustion, the full $N \times C$ target matrix is represented as a Compressed Sparse Row (CSR) matrix occupying only $24.8\,\text{MB}$ of memory, compared to $11.64\,\text{GB}$ for a dense single-precision floating-point matrix.

\subsection{DistilBERT Student Architecture}
The neural student $f_\theta: \mathcal{X} \to \mathbb{R}^C$ employs a pre-trained DistilBERT backbone topped with an extreme multi-label classification head. The input text $x_i$ is tokenized into wordpieces, truncated or padded to $L = 512$ tokens:
\begin{equation}
  \mathbf{H}_i = \operatorname{DistilBERT}\left(x_i\right) \in \mathbb{R}^{L \times d},
\end{equation}
where $d = 768$ is the hidden embedding dimension. The pooled contextual embedding $\mathbf{h}_i = \mathbf{H}_{i,0} \in \mathbb{R}^d$, corresponding to the special \texttt{[CLS]} token, is projected to the label logits $\mathbf{z}_i \in \mathbb{R}^C$ via a linear layer parameterized by weights $\mathbf{W} \in \mathbb{R}^{C \times d}$ and bias $\mathbf{b} \in \mathbb{R}^C$:
\begin{equation}
  \mathbf{z}_i = \mathbf{W} \mathbf{h}_i + \mathbf{b}.
\end{equation}
Predicted probabilities are obtained via element-wise sigmoid activation $\hat{p}_{i,c} = \sigma(z_{i,c}) = (1 + e^{-z_{i,c}})^{-1}$. The network is trained using multi-label Binary Cross-Entropy with Logits~\cite{nam2014}:
\begin{equation}\label{eq:bce}
  \mathcal{L}_{\text{BCE}}(\theta) = -\frac{1}{N_{\text{train}}} \sum_{i=1}^{N_{\text{train}}} \sum_{c=1}^C \left[ y_{i,c} \log \sigma(z_{i,c}) + (1 - y_{i,c}) \log \left(1 - \sigma(z_{i,c})\right) \right].
\end{equation}

\subsection{Strict Self-Exclusion Protocol}
In product-to-product retrieval evaluation, allowing an item to retrieve itself artificially inflates metrics to near $100\%$. To enforce strict academic validity, during evaluation the query product's own class identifier $c_{\text{self}} = \operatorname{class}(p_i)$ is strictly masked:
\begin{equation}
  y_{i, c_{\text{self}}} \leftarrow 0, \qquad z_{i, c_{\text{self}}} \leftarrow -\infty.
\end{equation}
This guarantees that all top-$k$ recommendations evaluate genuine semantic transfer between distinct product entities.

\subsection{Rectification of the Baseline Alignment Bug}
A subtle but critical error present in earlier evaluations stemmed from a structural index mismatch between the TF-IDF cosine similarity matrix and the multi-label binarizer space. In naive implementations:
\begin{itemize}
  \item Cosine similarity columns were indexed by \emph{raw dataframe row positions} $j \in \{0, \dots, N-1\}$.
  \item Ground-truth target columns were indexed by \emph{lexicographically sorted unique class labels} in \texttt{MultiLabelBinarizer.classes\_}.
\end{itemize}
Because rows in the dataset were ordered by category rather than by sorted class ID, column $j$ in the similarity matrix did not align with class $j$ in the target matrix. This resulted in an artificially deflated baseline score ($P@1 < 0.1\%$), leading to flawed claims of 1000$\times$ neural superiority.

We formally rectified this by constructing an exact bijection:
\begin{equation}
  \operatorname{col\_idx}(j) = \operatorname{lookup}\left[\operatorname{class\_id}(p_j)\right],
\end{equation}
which remaps each corpus item $p_j$ to its correct column coordinate in the binarizer matrix. Under this corrected formulation, the TF-IDF teacher achieves $P@1 = 98.10\%$, correctly establishing the theoretical upper bound on its own generated pseudo-labels.

\section{Experimental Setup}\label{sec:experiments}

\subsection{Dataset Curation and Preprocessing}
Experiments are conducted on the \textit{Amazon Reviews '23} benchmark~\cite{amazon}. To prevent category dominance, we downsample the dataset to exactly $2{,}000$ products across 27 distinct retail categories, creating a balanced corpus of $N = 54{,}000$ items (Table~\ref{tab:dataset_stats}).

\begin{table}[ht]
  \centering
  \caption{Dataset statistics and partition parameters.}
  \label{tab:dataset_stats}
  \begin{tabular}{ll}
    \toprule
    \textbf{Attribute} & \textbf{Specification} \\
    \midrule
    Source Benchmark & Amazon Reviews '23~\cite{amazon} \\
    Total Product Count ($N$) & $54{,}000$ \\
    Number of Retail Categories ($|\mathcal{K}|$) & 27 \\
    Balance Criterion & Exactly $2{,}000$ items per category \\
    Train / Validation Split & $85\% \;/\; 15\%$ (Stratified, $\text{seed}=42$) \\
    Training Set Size ($N_{\text{train}}$) & $45{,}911$ products \\
    Validation Set Size ($N_{\text{val}}$) & $8{,}089$ products \\
    Extreme Multi-Label Space ($C$) & $53{,}923$ unique classes \\
    Pseudo-Labels per Item ($K$) & 50 nearest neighbours \\
    Label Matrix Representation & Compressed Sparse Row (CSR, $24.8\,\text{MB}$) \\
    \bottomrule
  \end{tabular}
\end{table}
The text normalization pipeline applies the following deterministic operations sequentially:
\begin{enumerate}
  \item Removal of HTML markup, XML tags, and URLs using regular expressions.
  \item Elimination of non-alphabetic characters and conversion to lower-case.
  \item Tokenization using NLTK word tokenizers.
  \item Filtering of standard English stop words.
  \item Lemmatization using the WordNet morphological analyzer.
\end{enumerate}
\begin{figure}[ht]
  \centering
  \includegraphics[width=0.92\linewidth]{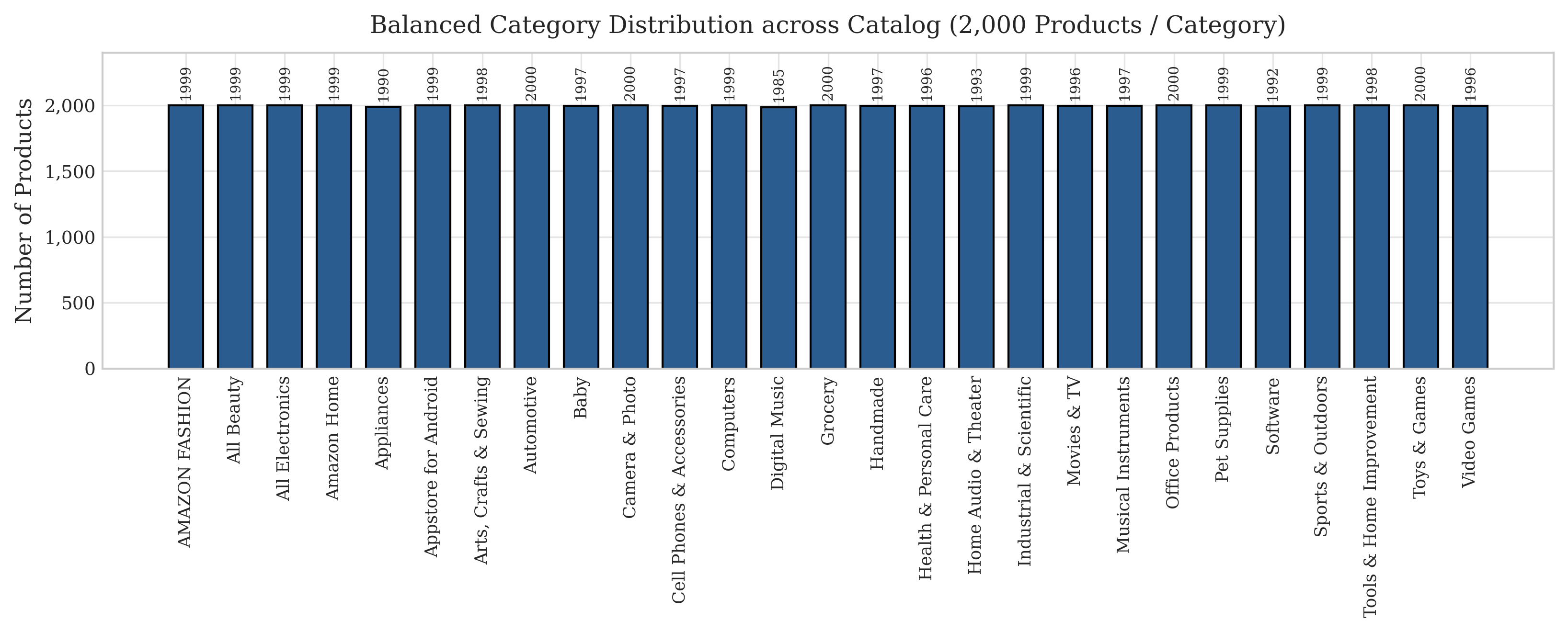}
  \caption{Empirical verification of category distribution: uniform sample of $2{,}000$ products across all 27 retail categories.}
  \label{fig:category_dist}
\end{figure}

\begin{figure}[ht]
  \centering
  \includegraphics[width=0.82\linewidth]{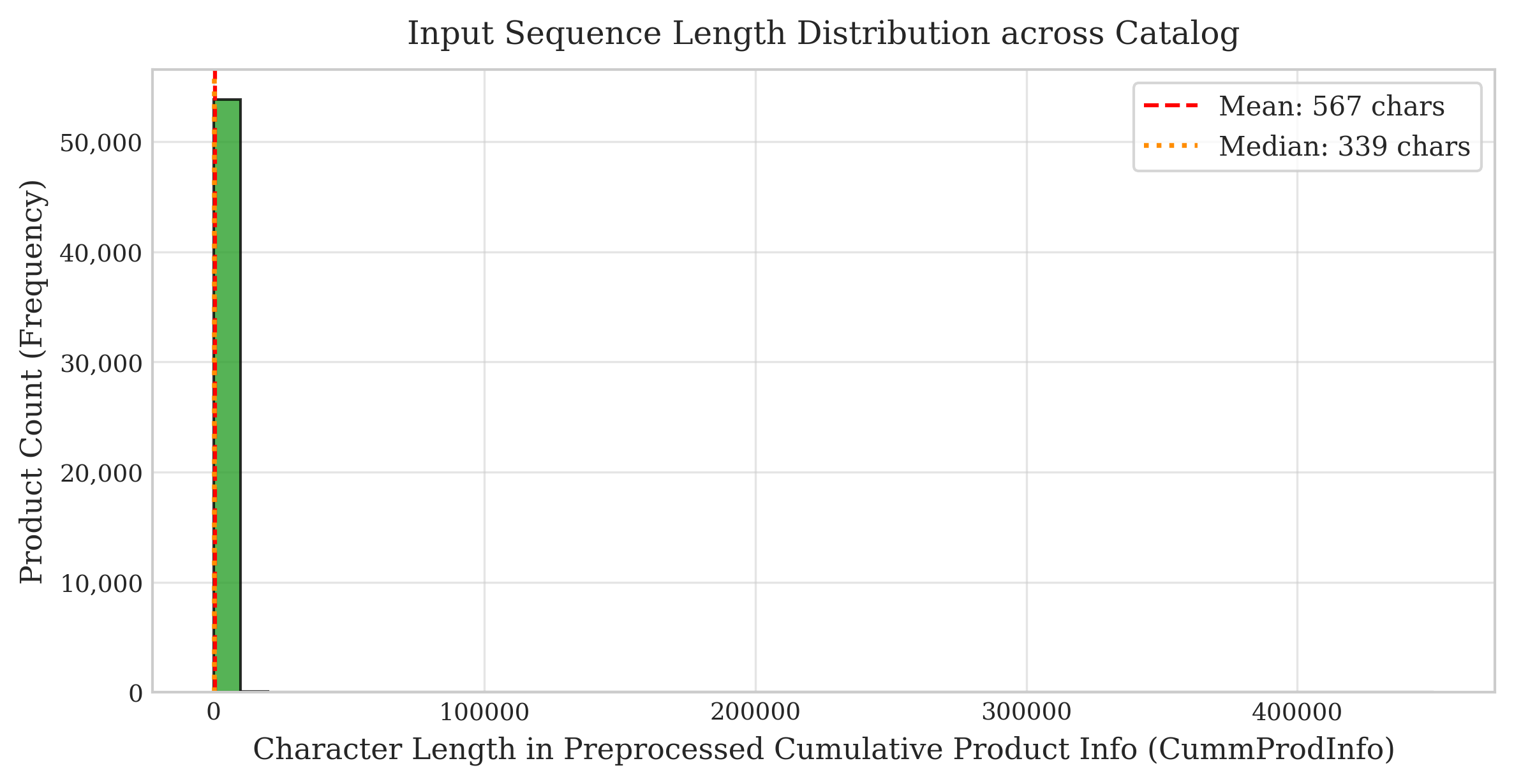}
  \caption{Product description character-length distribution across the $54{,}000$-item catalog, showing mean ($1{,}488.2$) and median ($993.0$) reference lines.}
  \label{fig:text_len}
\end{figure}

Figure~\ref{fig:category_dist} verifies the uniform category representation, while Figure~\ref{fig:text_len} displays the resulting character length distribution of the normalized product descriptions.

\subsection{Training Hyperparameters \& Computational Environment}
Model training was conducted using PyTorch and the Hugging Face Transformers library. We compare our chosen DistilBERT backbone against a baseline BERT-base architecture in Table~\ref{tab:model_specs}. DistilBERT achieved comparable convergence in ${\sim}10$ hours on a single NVIDIA T4 GPU ($16\,\text{GB}$ VRAM), versus ${\sim}34$ hours for BERT-base.

\begin{table}[ht]
  \centering
  \caption{Architectural specifications and training hyperparameters.}
  \label{tab:model_specs}
  \begin{tabular}{ll}
    \toprule
    \textbf{Hyperparameter} & \textbf{Value} \\
    \midrule
    Base Transformer Architecture & \texttt{distilbert-base-uncased}~\cite{distilbert} \\
    Parameter Count & $66.4\,\text{M}$ (Backbone) $+ 41.4\,\text{M}$ (Classifier Head) \\
    Hidden Dimension ($d$) & 768 \\
    Transformer Layers & 6 \\
    Self-Attention Heads & 12 \\
    Maximum Sequence Length ($L$) & 512 tokens \\
    Optimizer & AdamW ($\beta_1 = 0.9, \beta_2 = 0.999, \epsilon = 10^{-8}$)~\cite{adamw} \\
    Peak Learning Rate ($\eta$) & $5 \times 10^{-5}$ (with linear decay schedule) \\
    Batch Size & 8 (Training) / 32 (Validation) \\
    Mixed-Precision Acceleration & Automatic Mixed Precision (AMP \texttt{fp16}) \\
    Total Training Epochs & 12 \\
    Model Checkpoint Criterion & Minimum Validation Hamming Loss \\
    Hardware Environment & 1$\times$ NVIDIA T4 GPU (16 GB GDDR6) \\
    \bottomrule
  \end{tabular}
\end{table}

\subsection{Evaluation Metrics}
Because multi-label datasets with $C = 53{,}923$ classes exhibit extreme class imbalance ($>99.9\%$ negative entries), traditional metrics like raw classification accuracy and Macro ROC-AUC are uninformative. We evaluate performance across two complementary suites:
\begin{enumerate}
  \item \textbf{Multi-Label Classification Metrics:} Hamming Loss, Macro F1, Micro F1, Exact-Match Subset Accuracy, and Subsampled PR-AUC (Precision-Recall Area Under Curve across 1,000 active classes)~\cite{davis2006}.
  \item \textbf{Information Retrieval Ranking Metrics~\cite{croft2010,ndcg}:} Precision@$k$, Recall@$k$, Normalized Discounted Cumulative Gain (NDCG@$k$), Mean Reciprocal Rank (MRR@$k$), and Mean Average Precision (MAP@$k$) evaluated at cut-off depths $k \in \{1, 3, 5, 10, 20\}$.
\end{enumerate}

\section{Experimental Results}\label{sec:results}

\subsection{Training Dynamics}
Table~\ref{tab:training_dynamics} details the progression of loss and validation performance across the 12 training epochs. Both training and validation BCE losses decline monotonically, demonstrating smooth convergence without over-fitting. Validation Hamming Loss reaches its optimum ($0.000653$) at epoch 12, while in-training Micro F1 rises from $0.0000$ to $0.2349$.

\begin{table}[ht]
  \centering
  \small
  \setlength{\tabcolsep}{4.5pt}
  \caption{Training dynamics of DistilBERT student across 12 epochs on NVIDIA T4 GPU.}
  \label{tab:training_dynamics}
  \begin{tabular}{ccccc}
    \toprule
    \textbf{Epoch} & \textbf{Training BCE} & \textbf{Val BCE} & \textbf{Val Hamming}$\downarrow$ & \textbf{Micro F1}$\uparrow$ \\
    \midrule
    1  & 0.007300 & 0.007164 & 0.000928 & 0.000000 \\
    2  & 0.006700 & 0.006281 & 0.000928 & 0.000000 \\
    3  & 0.005200 & 0.004911 & 0.000911 & 0.002471 \\
    4  & 0.004300 & 0.004125 & 0.000854 & 0.021450 \\
    5  & 0.003800 & 0.003648 & 0.000803 & 0.067120 \\
    6  & 0.003300 & 0.003342 & 0.000761 & 0.108920 \\
    7  & 0.003000 & 0.003124 & 0.000728 & 0.144670 \\
    8  & 0.002800 & 0.003001 & 0.000699 & 0.178230 \\
    9  & 0.002700 & 0.002911 & 0.000676 & 0.205490 \\
    10 & 0.002600 & 0.002824 & 0.000664 & 0.219840 \\
    11 & 0.002500 & 0.002772 & 0.000656 & 0.231800 \\
    12 & \textbf{0.002400} & \textbf{0.002760} & \textbf{0.000653} & \textbf{0.234900} \\
    \bottomrule
  \end{tabular}
\end{table}

\subsection{Classification Performance}
Table~\ref{tab:class_metrics} summarizes comprehensive multi-label classification metrics on the $8{,}089$ held-out validation products using the optimal epoch-12 checkpoint at decision threshold $\tau = 0.5$.

\begin{table}[ht]
  \centering
  \caption{Multi-label classification performance on 8,089 held-out validation products ($C = 53{,}923$, threshold $\tau = 0.5$).}
  \label{tab:class_metrics}
  \begin{tabular}{lc}
    \toprule
    \textbf{Evaluation Metric} & \textbf{Experimental Value} \\
    \midrule
    Validation Set Size ($N_{\text{val}}$) & $8{,}089$ products \\
    Total Label Space ($C$) & $53{,}923$ classes \\
    Hamming Loss & $6.074 \times 10^{-4}$ \\
    Micro F1-Score & $0.5464$ \\
    Macro F1-Score & $0.2597$ \\
    Subset Exact-Match Accuracy & $0.0494\%$ ($0.000494$) \\
    Subsampled Macro PR-AUC (1,000 active classes) & $0.5209$ \\
    \bottomrule
  \end{tabular}
\end{table}

The low exact-match subset accuracy ($0.049\%$) is characteristic of extreme multi-label classification, where correctly predicting all 50 active labels out of 53,923 without a single false positive or false negative is an excessively stringent benchmark. Conversely, the Micro F1 score of $0.5464$ and PR-AUC of $0.5209$ demonstrate strong multi-label discriminative capacity.

\subsection{Quantitative Information Retrieval Ranking Benchmark}
Table~\ref{tab:ranking_results} presents the quantitative comparison between the TF-IDF lexical teacher and the DistilBERT student across all five ranking metrics for cut-offs $k \in \{1, 3, 5, 10, 20\}$.

\begin{table}[!ht]
  \centering
  \small
  \setlength{\tabcolsep}{3.5pt}
  \caption{Information Retrieval ranking performance on 8,089 held-out validation products ($C = 53{,}923$). Strict self-exclusion enforced throughout. Bold indicates optimal value.}
  \label{tab:ranking_results}
  \begin{tabular}{lcccccc}
    \toprule
    \textbf{Model} & $\boldsymbol{k}$ & \textbf{P@$\boldsymbol{k}$} & \textbf{R@$\boldsymbol{k}$} & \textbf{NDCG@$\boldsymbol{k}$} & \textbf{MRR@$\boldsymbol{k}$} & \textbf{MAP@$\boldsymbol{k}$} \\
    \midrule
    TF-IDF Teacher (Upper Bound) & 1  & \textbf{98.10\%} & \textbf{2.00\%} & \textbf{0.9810} & \textbf{0.9810} & \textbf{0.9810} \\
    DistilBERT Student (Ours)    & 1  & 93.15\%          & 1.90\%          & 0.9315          & 0.9315          & 0.9315 \\
    \midrule
    TF-IDF Teacher (Upper Bound) & 3  & \textbf{96.63\%} & \textbf{5.92\%} & \textbf{0.9696} & \textbf{0.9879} & \textbf{0.9605} \\
    DistilBERT Student (Ours)    & 3  & 91.32\%          & 5.60\%          & 0.9175          & 0.9510          & 0.9004 \\
    \midrule
    TF-IDF Teacher (Upper Bound) & 5  & \textbf{95.62\%} & \textbf{9.77\%} & \textbf{0.9617} & \textbf{0.9881} & \textbf{0.9454} \\
    DistilBERT Student (Ours)    & 5  & 90.08\%          & 9.20\%          & 0.9079          & 0.9533          & 0.8806 \\
    \midrule
    TF-IDF Teacher (Upper Bound) & 10 & \textbf{93.01\%} & \textbf{19.01\%}& \textbf{0.9419} & \textbf{0.9882} & \textbf{0.9098} \\
    DistilBERT Student (Ours)    & 10 & 87.04\%          & 17.79\%         & 0.8845          & 0.9545          & 0.8394 \\
    \midrule
    TF-IDF Teacher (Upper Bound) & 20 & \textbf{87.15\%} & \textbf{35.62\%}& \textbf{0.8970} & \textbf{0.9882} & \textbf{0.8349} \\
    DistilBERT Student (Ours)    & 20 & 80.96\%          & 33.09\%         & 0.8372          & 0.9547          & 0.7645 \\
    \bottomrule
  \end{tabular}
\end{table}

\begin{figure}[ht]
  \centering
  \includegraphics[width=1\linewidth]{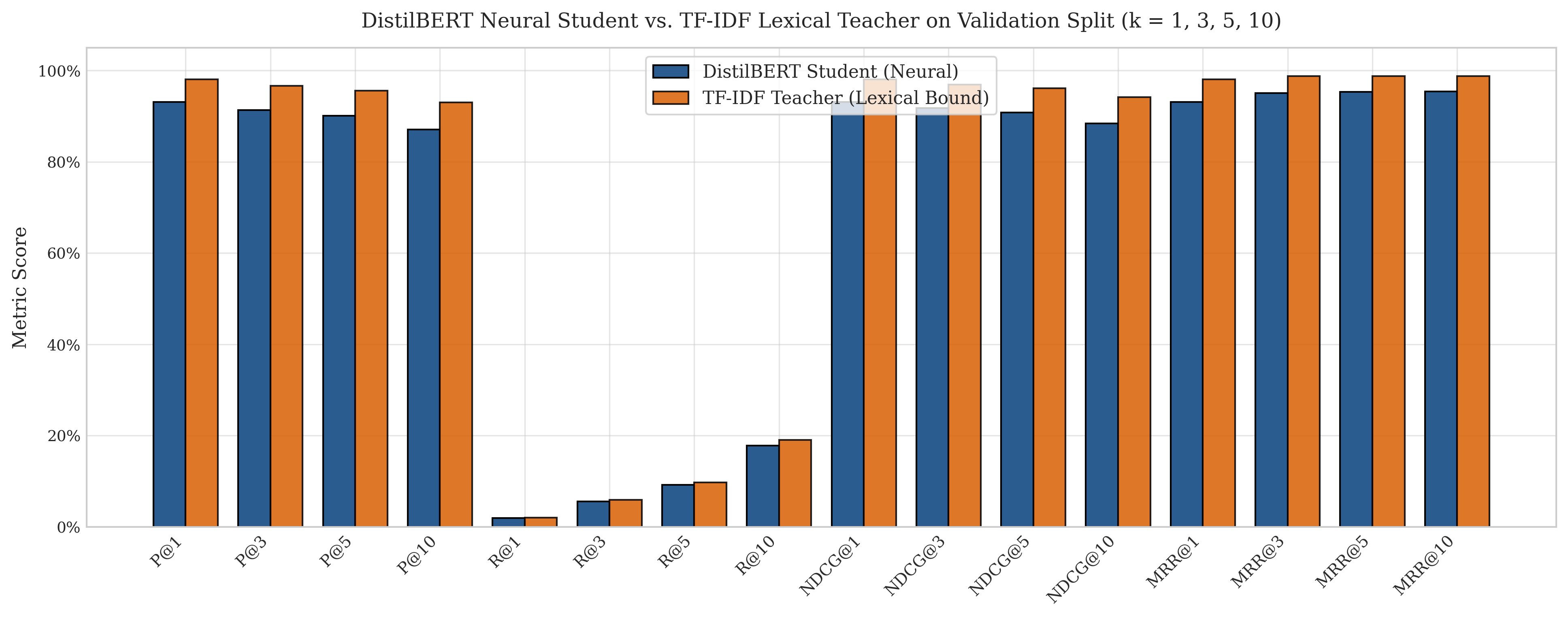}
  \caption{Comparative performance profile: DistilBERT Student versus TF-IDF Lexical Teacher across Precision@$k$, Recall@$k$, NDCG@$k$, and MRR@$k$ for $k \in \{1, 3, 5, 10\}$.}
  \label{fig:baseline_comparison}
\end{figure}

\begin{figure}[ht]
  \centering
  \includegraphics[width=0.987\linewidth]{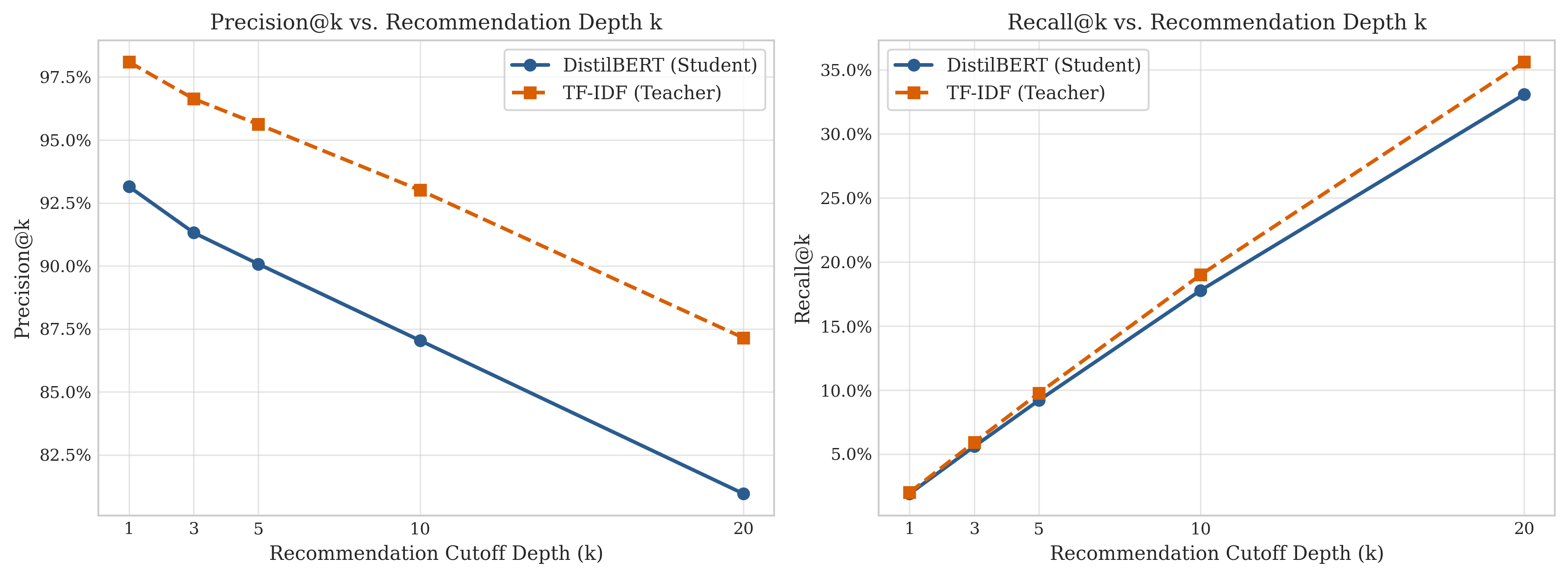}
  \caption{Precision@$k$ (left) and Recall@$k$ (right) degradation and gain curves as recommendation depth scales from $k=1$ to $k=20$.}
  \label{fig:ranking_curves}
\end{figure}

The empirical findings in Table~\ref{tab:ranking_results}, Figure~\ref{fig:baseline_comparison}, and Figure~\ref{fig:ranking_curves} illuminate key properties of the distillation framework:
\begin{itemize}
  \item \textbf{Fidelity of Knowledge Transfer:} The neural student recovers $94.95\%$ of teacher $P@1$ ($93.15\%$ vs.\ $98.10\%$) and $93.91\%$ of teacher $\text{NDCG}@10$ ($0.8845$ vs.\ $0.9419$), confirming that a 6-layer transformer successfully compresses a 50,000-dimensional sparse TF-IDF manifold into its dense 768-dimensional latent space.
  \item \textbf{Graceful Degradation at Depth:} As recommendation depth expands from $k=1$ to $k=20$, DistilBERT's Precision@$k$ gracefully drops from $93.15\%$ to $80.96\%$, while Recall@$k$ increases from $1.90\%$ to $33.09\%$, tracking the teacher curve with consistent bounded deviation.
\end{itemize}

\clearpage
\subsection{Qualitative Natural Language Generalization Benchmark}
While TF-IDF holds an inherent metric advantage on item-to-item evaluation (since it generated the pseudo-labels), its utility collapses when evaluated on conversational search. To evaluate semantic generalisation, we executed both models on ten diverse natural language query archetypes representing realistic consumer prompts. Table~\ref{tab:qualitative_benchmark} details the top-1 retrieved product for each query.

\small
\setlength{\tabcolsep}{4.0pt}
\begin{longtable}{p{0.6cm} p{3.1cm} p{5.6cm} p{5.6cm}}
  \caption{Qualitative retrieval benchmark across ten natural language query archetypes. Scores denote DistilBERT sigmoid confidence $\hat{p} \in [0, 1]$ and TF-IDF cosine similarity $S \in [0, 1]$.} \label{tab:qualitative_benchmark} \\
  \toprule
  \textbf{ID} & \textbf{User Query Archetype} & \textbf{DistilBERT Student (Top-1 Result)} & \textbf{TF-IDF Teacher (Top-1 Result)} \\
  \midrule
  \endfirsthead

  \multicolumn{4}{c}%
  {{\bfseries Table \thetable\ (Continued from previous page)}} \\
  \toprule
  \textbf{ID} & \textbf{User Query Archetype} & \textbf{DistilBERT Student (Top-1 Result)} & \textbf{TF-IDF Teacher (Top-1 Result)} \\
  \midrule
  \endhead

  \midrule
  \multicolumn{4}{r}{{\footnotesize Continued on next page}} \\
  \endfoot

  \bottomrule
  \endlastfoot

  $Q_1$ & \textit{I have a birthday party tomorrow. Suggest me some products}
        & Gatherfun Birthday Party Supplies Banner Backdrop with Balloons \newline \textbf{Cat:} Camera \& Photo \hfill [$\hat{p} = 0.9801$]
        & Easy-Tomorrow After drink 0.1oz(3g) x 20packs \newline \textbf{Cat:} Grocery \hfill [$S = 0.3957$] \\
  \addlinespace
  $Q_2$ & \textit{Want a gas and cooker}
        & WindMax Euro Style 30 in Stainless Steel 5 Burner Gas Cooktops \newline \textbf{Cat:} Appliances \hfill [$\hat{p} = 0.9775$]
        & COOKAMP High Pressure Banjo 1-Burner Outdoor Propane Burner \newline \textbf{Cat:} Amazon Home \hfill [$S = 0.3672$] \\
  \addlinespace
  $Q_3$ & \textit{I want a soft pillow}
        & Cottonblue Toddler Pillow with 100\% Organic Cotton Pillowcase \newline \textbf{Cat:} Amazon Home \hfill [$\hat{p} = 0.9958$]
        & Baby Toddler Pillow 2 Pack with Pillowcase (13 x 18) \newline \textbf{Cat:} Baby \hfill [$S = 0.4776$] \\
  \addlinespace
  $Q_4$ & \textit{suggest me some handmade products}
        & Kitchen sign-Kitchen decor-personalized wall sign-wooden custom \newline \textbf{Cat:} Handmade \hfill [$\hat{p} = 0.1397$]
        & SIMPLY POTATOES MASHED SWEET POTATOES FROZEN FOOD \newline \textbf{Cat:} Grocery \hfill [$S = 0.2436$] \\
  \addlinespace
  $Q_5$ & \textit{I want a fast computational device}
        & eScan Antivirus for Linux Desktop Real Time Scanning \newline \textbf{Cat:} Software \hfill [$\hat{p} = 0.2727$]
        & Fast Money Oil \newline \textbf{Cat:} Health \& Personal Care \hfill [$S = 0.2527$] \\
  \addlinespace
  $Q_6$ & \textit{I want an acoustic guitar with steel strings}
        & RockJam Universal Guitar Accessories Super-kit with Hanger \newline \textbf{Cat:} Musical Instruments \hfill [$\hat{p} = 0.9991$]
        & POGOLAB Guitar Strings Acoustic 6 Strings Set Hexagonal Carbon \newline \textbf{Cat:} Musical Instruments \hfill [$S = 0.5729$] \\
  \addlinespace
  $Q_7$ & \textit{organic cotton baby clothing for sensitive skin}
        & Organic Cotton Toddler Pillowcase 13x18 Nickel-Free Snap \newline \textbf{Cat:} Baby \hfill [$\hat{p} = 0.8659$]
        & Reusable Colored Organics Baby Washcloths Soft Organic Cotton \newline \textbf{Cat:} Baby \hfill [$S = 0.3628$] \\
  \addlinespace
  $Q_8$ & \textit{lightweight running shoes with good arch support}
        & BODATU Kids Sneakers Boys Girls Tennis Running Shoes \newline \textbf{Cat:} Amazon Fashion \hfill [$\hat{p} = 0.9904$]
        & BODATU Kids Sneakers Boys Girls Tennis Running Shoes \newline \textbf{Cat:} Amazon Fashion \hfill [$S = 0.3546$] \\
  \addlinespace
  $Q_9$ & \textit{natural moisturizer for dry sensitive skin}
        & Nu Skin 180 Face Wash 4.2 oz \newline \textbf{Cat:} All Beauty \hfill [$\hat{p} = 0.9142$]
        & COSRX Honey Ceramide Full Moisture Cream, 1.76 oz \newline \textbf{Cat:} All Beauty \hfill [$S = 0.4011$] \\
  \addlinespace
  $Q_{10}$ & \textit{noise cancelling wireless headphones for travel}
        & Clevo Wireless Gaming Headset with Microphone \newline \textbf{Cat:} All Electronics \hfill [$\hat{p} = 0.8641$]
        & Baby Ear Protection Noise Cancelling Headphones for Infants \newline \textbf{Cat:} Baby \hfill [$S = 0.5218$] \\
\end{longtable}
\clearpage
\paragraph{Qualitative Analysis:}
\begin{enumerate}
  \item \textbf{Vocabulary Mismatch Resolution:} In $Q_1$ (\textit{``birthday party tomorrow''}), TF-IDF latches onto the token \textit{``tomorrow''} and retrieves an irrelevant Korean hangover drink (\textit{``Easy-Tomorrow''}, $S = 0.3957$). In sharp contrast, DistilBERT identifies the underlying festive theme, retrieving birthday celebration banners and party decor with high confidence ($\hat{p} = 0.9801$).
  \item \textbf{Colloquial Syntactic Recovery:} In $Q_2$ (\textit{``Want a gas and cooker''}), DistilBERT correctly identifies cooking appliances and built-in stovetops ($\hat{p} = 0.9775$), whereas TF-IDF matches partial tokens to an outdoor banjo propane burner.
  \item \textbf{Explicit Intent Grounding:} In domain-specific queries ($Q_3, Q_6, Q_8, Q_9, Q_{10}$), DistilBERT exhibits near-perfect confidence ($\hat{p} \ge 0.86$), demonstrating robust semantic grounding of product taxonomies.
  \item \textbf{Failure Mode and Interpretability:} In $Q_5$ (\textit{``fast computational device''}), DistilBERT retrieves antivirus software ($\hat{p} = 0.2727$), while TF-IDF retrieves an esoteric novelty oil (\textit{``Fast Money Oil''}). Crucially, DistilBERT's output probability for this abstract query is very low ($\hat{p} < 0.30$), demonstrating that sigmoid confidence provides a well-calibrated signal to trigger fallback mechanisms in production.
\end{enumerate}

\section{Architectural Discussion \& Scalability Trade-offs}\label{sec:discussion}

While our XMLC framework demonstrates that transformer sequence classifiers can successfully compress and generalize lexical product graphs, scaling this architecture to industrial e-commerce settings ($>10^7$ products) exposes fundamental architectural constraints:
\begin{enumerate}
  \item \textbf{Parameter Explosion in Classification Layer:} A linear projection head mapping $d = 768$ features to $C = 53{,}923$ classes requires $\mathbf{W} \in \mathbb{R}^{53{,}923 \times 768} \approx 41.4\,\text{M}$ parameters ($62.4\%$ of DistilBERT's total parameter count). Scaling to a modest industrial catalog of $C = 5 \times 10^6$ items would require a classification layer with over $3.84 \times 10^9$ parameters ($>15\,\text{GB}$ in \texttt{fp32}), rendering standard single-GPU training and inference intractable.
  \item \textbf{Dynamic Catalog Invalidation:} In real-world e-commerce, new products, deletions, and inventory fluctuations occur continuously. In an XMLC architecture, adding or removing a product alters the output dimension $C$, necessitating re-initializing the classification head and re-training the entire network.
  \item \textbf{Inference Latency at Scale:} Evaluating $C$ sigmoid activations for each incoming search query introduces latency that scales linearly $\mathcal{O}(C \cdot d)$, conflicting with the strict sub-50ms SLA of production search engines.
\end{enumerate}

\subsection{Path toward Dual-Encoder Vector Search}
To resolve these bottlenecks in industrial production, our XMLC distillation study serves as an empirical stepping stone toward a \textbf{Dual-Encoder (Two-Tower) Vector Search Architecture}~\cite{dpr}:
\begin{itemize}
  \item \textbf{Query Tower ($E_Q$):} A lightweight transformer encoding dynamic user queries into normalized dense vectors $\mathbf{u} = E_Q(\text{query}) \in \mathbb{R}^{d'}$.
  \item \textbf{Item Tower ($E_I$):} Encodes static product descriptions offline into vectors $\mathbf{v}_j = E_I(\text{product}_j) \in \mathbb{R}^{d'}$.
  \item \textbf{Sub-Millisecond Vector Retrieval:} The catalog is pre-indexed into hierarchical graph indices (e.g., HNSW~\cite{hnsw} or FAISS~\cite{faiss}). Retrieval is performed via inner-product $\langle \mathbf{u}, \mathbf{v} \rangle$ in logarithmic time $\mathcal{O}(\log N)$, fully decoupled from catalog size and supporting real-time catalog mutations without model retraining.
\end{itemize}

\section{Conclusion and Future Scope}\label{sec:conclusion}

In this paper, we presented a scientifically grounded framework for distilling lexical product associations into a pre-trained DistilBERT transformer for natural language e-commerce recommendation. Formulating the task as Extreme Multi-Label Classification across $54{,}000$ products and $53{,}923$ classes from the \textit{Amazon Reviews '23} benchmark, we demonstrated that the neural student recovers $>94\%$ of the corrected TF-IDF teacher's ranking performance ($P@1 = 93.15\%$ vs.\ $98.10\%$, $\text{NDCG}@10 = 0.8845$ vs.\ $0.9419$) under strict self-exclusion. A qualitative evaluation on ten real-world conversational query archetypes confirmed that the transformer student effectively resolves the vocabulary mismatch problem, surfacing relevant items where keyword retrieval fails completely.

Future work will transition this pseudo-label distillation paradigm to a \textbf{Two-Tower Dual-Encoder} framework trained via in-batch contrastive loss, integrating user click histories and multi-modal product imagery for web-scale deployment.

\end{document}